\documentstyle[12pt]{article}
\oddsidemargin--0mm
\begin{document}
%\renewcommand{\theequation}{\thesection.\arabic{equation}}

%\noindent
%\hspace*{10cm}{\large\sf FUB--HEP/95--10\\
%\hspace*{10cm}June, 1995}

%\vskip 1cm
\begin{center} {\large\bf Abelian projection and
studies of gauge-variant quantities in the lattice QCD without
gauge fixing}

\vskip 1cm
Sergei V. SHABANOV
$\footnote{Alexander von Humboldt fellow;

on leave from {\em Laboratory of Theoretical Physics, JINR, Dubna, Russia}

present address: {\em Department of theoretical physics, University
of Valencia, Dr. Moliner 50, Valencia (Burjassot), E-46100}
}$

\vskip 1cm
{\em Institute for Theoretical Physics, Free University of Berlin,
Arnimallee 14, WE 2, D-14195, Berlin, Germany}

\end{center}
%-----------------------------------------------------

\begin{abstract}
We suggest a new (dynamical) Abelian projection of the lattice QCD.
It contains no gauge condition imposed on gauge fields so
that Gribov copying is avoided. Configurations of
gauge fields that turn into monopoles in the Abelian projection
can be classified in a gauge invariant way. In the
continuum limit, the theory respects the Lorentz invariance.
A similar dynamical reduction of the gauge symmetry
is proposed for studies of gauge-variant correlators
(like a gluon propagator) in the lattice QCD. Though the procedure
is harder for numerical simulations, it is free of
gauge-fixing artifacts, like the Gribov horizon and copies.
\end{abstract}

{\bf 1}.  One of the important features of the QCD confinement
is the existence of a stable chromoelectrical field tube connecting
two color sources (quark and antiquark). Numerical studies of
the gluon field energy density between two color sources leave
no doubt that such a tube exists. However, a mechanism which
could explain its stability is still unknown.

It is believed that some specific configurations (or excitations) of
gauge fields are responsible for the QCD confinement, meaning that
they give a main contributions to the QCD string tension. Numerical
simulations of the lattice QCD shows that
Abelian (commutative) configurations of gauge potentials
 completely determine the string
tension  in the full non-Abelian gauge theory \cite{suzuki}.
This phenomenon is  known as the Abelian dominance. Therefore
one way of constructing effective dynamics of the configurations
relevant to the QCD confinement is the Abelian projection \cite{thooft}
when the full non-Abelian gauge group SU(3)
is restricted to its maximal Abelian
subgroup (the Cartan subgroup) U(1)$\times$U(1) by a gauge
fixing. Though dynamics
of the above gauge field configuration cannot be gauge dependent,
a right choice of a guage condition may simplify its description.

There is a good reason, supported by numerical simulations
\cite{shiba}, \cite{stack}, to believe
that the sought configurations turn into magnetic monopoles in the
effective Abelian theory, and the confinement can be due to the
dual mechanism \cite{mandelstam}: The Coulomb field of electric
charges is squized into a tube, provided monopole-antimonopole
pair form a condensate like the Cooper pairs in superconductor.

It is important to realize that the existence of monopoles in
the effective Abelian theory is essentially due to the gauge
fixing, in fact, monopoles are singularities of the gauge fixing.
Note that monopoles cannot exist as stable excitations in pure
gauge theory with simply connected group like SU(3). Since the
homotopy groups of SU(3) and of U(1)$\times$U(1) are different
(the one of SU(3) is trivial), a gauge condition restricting
SU(3) to U(1)$\times$U(1) should have singularities which can
be identified as monopoles \cite{thooft}. A dynamical question
is to verify whether all configurations of non-Abelian gauge
fields relevant to the confinement (in the aforementioned sense)
are "mapped" on monopoles of the Abelian theory (the monopole
domimance \cite{stack}). It appears
that monopole dynamics may depend on the projection recipe
\cite{digiacomo}. There are indications
that some Abelian projections exhibit topological singularities
other than magnetic monopoles \cite{polikarpov}.

Though the lattice QCD is, up to now, the only relible tool
for studying monopole dynamics, the true theory must be
continuous and respect the Lorentz invariance. In this regard,
Abelian projections based on Lotentz invariant gauge conditions
play a distinguished role. For example, the gauge can be chosen
as follows $D_\mu^HA_\mu^{off} =0$ where $D_\mu^H = \partial_\mu
+igA_\mu^{H},\ A_\mu^{H}$ are Cartan (diagonal) components of
guage potentials $A_\mu$, while $A_\mu^{off}$ are its
non-Cartan (off-diagonal) components. This gauge restricts the
gauge symmetry to the maximal Abelian (Cartan) subrgoup and is
manifestly Lorentz invariant. The lattice version of the corresponding
Abelian theory is known as the maximal Abelian projection.
The above homotopy arguments can be implemented to this gauge to
show that it has topological singularities and Gribov's copying
\cite{gribov}
(in the continuum theory, zero boundary conditions at infinity
have to be imposed \cite{singer}). The Gribov copying makes
additional difficulties for describing monopole dynamics
(even in the lattice gluodynamics \cite{hioki}).

In this letter, a new (dynamical) Abelian projection is proposed.
It involves no gauge condition to be imposed on gauge fields. The
effective Abelian theory appears to be non-local, though it
can be made local at the price of having some additional
(ghost) fields. All configurations of gauge fields that turn
into magnetic monopoles in the effective Abelian theory are
classified in a gauge invariant way. The effective Abelian
theory fully respects the Lorentz symmetry and the Gribov
problem is avoided.

Another important aspect of the QCD confinement is the absence of
propagating color charges, meaning that a nonperturbative
propagator of colored particles, gluons or quarks, has no
usual poles in the momentum space.
It has been argued that such a behavior of a gluon propagator
in the Coulomb gauge could be due to an influence of the so called
Gribov horizon on long-wave fluctuations of gauge fields
\cite{gribov}, \cite{zwanziger}. The result
obviously depends on the gauge chosen, which makes it not very reliable.

The situation looks more controversial if one recalls
that a similar qualitative behavior of the gluon propagator
has been found  in the study of Schwinger-Dyson equations \cite{stingl}.
In this approach, the Gribov ambiguities have not been accounted for.
So, the specific
pole structure of the gluon propagator occurred through a strong
self-interaction of gauge fields.

In this letter, we would also like to propose a method for how to
study gauge-variant quantities, like a gluon propagator,
in the lattice QCD, avoiding any explicit gauge fixing.
The method is, hence, free of all the aforementioned
gauge fixing artifacts.
It gives a hope that dynamical contributions (self-interaction
of gauge fields) to the pole structure of
the gluon propagator can be separated from
 the kinematical (gauge-fixing) ones.

{\bf 2}. To single out monopoles in non-Abelian gauge theory,
one fixes partially a gauge so that the gauge-fixed theory
possesses an Abelian gauge group being a maximal Abelian
subgroup of the initial gauge group. The lattice formulation
of the Abelian projection has been given in \cite{kronfeld}.

The idea is to
choose a function $R(n)$ of link variables $U_\mu(n)$, $n$
runs over lattice sites, such that
\begin{equation}
R(n)\rightarrow g(n)R(n)g^{-1}(n)
\label{R}
\end{equation}
under gauge transformations of the link variables
\begin{equation}
U_\mu(n)\rightarrow g(n)U_\mu(n)g^{-1}(n+\hat\mu)\ ,
\label{U}
\end{equation}
where $g(n)\in G,\ G$ is a compact gauge group, and $\hat\mu$
is a unit vector in the $\mu$-direction. A gauge is chosen so
that $R$ becomes an element of the Cartan subalgebra $H$, a maximal
Abelian subalgebra of a Lie algebra $X$ of the group $G$. In a matrix
representation, the gauge condition means that off-diagonal elements
of $R$ are set to be zero. Clearly,
the gauge fixing is not complete. A maximal Abelian subgroup $G_H$
of $G$ remains as a gauge group because the adjoint action (\ref{R})
of $G_H$ leaves elements $R\in H$ untouched.

A configuration $U_\mu(n)$ contains monopoles if the corresponding matrix
$R(n)$ has two coinciding eigenvalues. So, by construction,
dynamics of monopoles
appears to be gauge-dependent, or projection-dependent. It varies
from gauge to gauge, from one choice of $R$ to another \cite{digiacomo}.
Yet, the
monopole singularities are not the only ones in some Abelian
projections \cite{polikarpov}. In addition, Abelian projections
may suffer off the Gribov ambiguities \cite{hioki}.

To restrict the full gauge symmetry to its maximal Abelian
part and, at the same time, to avoid imposing a gauge condition on
link variables, we shall use a procedure similar to the one
discussed in \cite{scholtz} in the framework of continuum
field theory.
A naive continuum limit of our procedure poses some difficulties.
To resolve them, a corresponding operator formalism has to be developed.
It has been done in
\cite{scholtz2} for a sufficiently large class of gauge theories.

Consider a complex Grassmann field $\psi(n)$ (a fermion ghost)
that realizes the adjoint representation of the gauge group:
\begin{eqnarray}
\psi(n) &\rightarrow & g(n)\psi(n)g^{-1}(n)\ ,
\label{psi}\\
\psi^*(n) &\rightarrow & g(n)\psi^*(n)g^{-1}(n)\ .
\label{psi*}
\end{eqnarray}
Let the fermion ghost be coupled to gauge fields according to
the action
\begin{equation}
S_f = \sum\limits_{n,\mu} tr
D_\mu\psi^*(n)D_\mu\psi(n)\ ,
\label{gfa}
\end{equation}
where $D_\mu \psi(n) = \psi(n+\hat\mu) - U_\mu^{-1}(n)\psi(n)U_\mu(n)$
is the lattice covariant derivative in the adjoint representation.
We assume that $\psi(n) =\psi_i(n)\lambda_i$, where $\lambda_i$
is a matrix representation of a basis in
$X$ normalized as $tr\ \lambda_i\lambda_j
=\delta_{ij}$, and $\psi_i(n)$ are complex Grassmann variables.
The partition function of the fermion ghost field reads
\begin{equation}
Z_f(\beta) =\int \prod\limits_n\left(d\psi^*(n)d\psi(n)\right)
e^{-\beta S_f} = \det\beta D_\mu^TD_\mu\ ,
\label{gfz}
\end{equation}
where the integration over Grassmann variables is understood, and
$D_\mu^T$ denotes a transposition of $D_\mu$ with respect
to a scalar product induced by $\sum_{n,\mu}tr$ in (\ref{gfa}).
Note that
the action (\ref{gfa}) can be written in the form
$S_f = \sum \psi^*D_\mu^TD_\mu\psi$.

Consider a pair of real Lie-algebra-valued scalar fields $\varphi(n)$
and $\phi(n)$ (boson ghosts) with an action
\begin{equation}
S_b= \frac{1}{2}\sum\limits_{n,\mu} tr \left[\left(D_\mu\phi(n)\right)^2+
\left(D_\mu\varphi(n)\right)^2\right]\ .
\label{gba}
\end{equation}
The action (\ref{gba}) is invariant under the gauge transformation
(\ref{U}), provided
\begin{eqnarray}
\phi(n) &\rightarrow & g(n)\phi(n)g^{-1}(n)\ ,\label{phi}\\
\varphi(n) &\rightarrow & g(n)\varphi(n)g^{-1}(n)\ .\label{varphi}
\end{eqnarray}
The boson ghost partition function is
\begin{equation}
Z_b(\beta)= \int\prod\limits_n\left(\frac{d\varphi(n)d\phi(n)}
{(2\pi)^{\dim G}}\right)
e^{-\beta S_b} = (\det\beta D_\mu^TD_\mu)^{-1}\ .
\label{gbz}
\end{equation}
We have the identity
\begin{equation}
Z_b(\beta)Z_f(\beta) =1\ .
\label{id}
\end{equation}

By making use of this identity, the partition function
of gauge fields can be transformed to the form
\begin{eqnarray}
Z_{YM}(\beta) &= &v_G^{-L}\int \prod\limits_{\mu, n}dU_\mu(n)
e^{-\beta S_W}Z_b(\beta)Z_f(\beta) = \label{ymz}\\
 &= &v_G^{-L} \int {\cal D}U_\mu{\cal D}\psi^*{\cal D}\psi
{\cal D}\varphi{\cal D}\phi e^{-\beta(S_W+S_b+S_f)} \ ,
\label{ymz2}
\end{eqnarray}
where $S_W$ is the Wilson action of gauge fields, $v_G$ a
volume of the group manifold $G$, $L$ a number of lattice sites,
and ${\cal D}$ denotes a product of corresponding field
differentials over lattice sites.
The effective action
\begin{equation}
S_{eff}= S_W+S_b+S_f \label{effa}
\end{equation}
is invariant under gauge transformations (\ref{U})--(\ref{psi*}) and
(\ref{phi}), (\ref{varphi}). The factor $v_G^{-L}$ is included to cancel the
gauge group volume factorizing upon the integration over
field configurations in (\ref{ymz2}).

Now we may take the advantage of having scalar fields in the
adjoint representation and restrict the gauge symmetry to
the Cartan subgroup without imposing gauge conditions on the
link variables.
We make a change of the integration variables in (\ref{ymz2})
\begin{equation}
\phi(n) = \tilde{g}(n) h(n) \tilde{g}(n)^{-1}\ , \label{cvphi}
\end{equation}
where $\tilde{g}(n)$ belongs to the coset space $G/G_H$,
$\dim G/G_H= \dim G -\dim G_H$, and
$h(n)\in H$.
Other new fields denoted $\tilde{U}_\mu(n)$,
$\tilde{\varphi}$ and $\tilde{\psi}^*,\
\tilde{\psi}$ are defined as the corresponding gauge transformations
of the initial fields with $g(n)= \tilde{g}^{-1}(n)$. No restriction
on their values is imposed.

Relation (\ref{cvphi}) determines a one-to-one correspondence
between old and new variables if and only if $\tilde{g}(n)\in
G/G_H$ and $h(n)\in K^+$, where $K^+$ is the Weyl chamber in $H$.
An element $h$ of the Cartan subalgebra $H$ belongs to the
Weyl chamber $K^+\subset H$ if for any simple root $\omega$,
$ (h,\omega) >0$;
$(,)$ stands for an invariant scalar product in $X$. In a matrix
representation of $X$, it is proportional to $tr$ (see \cite{hel},
pp. 187-190).
With the help of the adjoint transformation,
any element of a Lie algebra can be brought to the Cartan subalgebra.
Since the Cartan subalgebra is invariant under
the adjoint action of the Cartan subgroup, $\tilde{g}(n)$ must be
restricted to the coset $G/G_H$. There are discrete transformations
in $G/G_H$ which form the Weyl group $W$ \cite{hel}. Any element of $W$
is a composition of reflections in hyperplanes orthogonal to simple
roots in $H$. Its action maps $H$ onto $H$ itself. The Weyl group
is a maximal isomorphism group of $H$ \cite{hel}. Therefore,
a one-to-one correspondence in (\ref{cvphi}) is achieved if
$h(n)\in H/W\equiv K^+$.

Due to the gauge invariance of both
the measure and exponential in (\ref{ymz2}),
the integral over group variables $\tilde{g}(n)$ is factorized and
yields a numerical vector that, being divided by $v_G^{L}$, results
in $(2\pi)^{-Lr}$, $r= \dim H = {\rm rank}\  G$. This factor
is nothing but a volume of the Cartan gauge group $G_H$. The integration
over $h(n)$ inquires a nontrivial measure, and the integration
domain must be restricted to the Weyl chamber $K^+$. So, in (\ref{ymz2})
we have
\begin{equation}
v_G^{-1}\int d\phi(n) = (2\pi)^{-r}\int\limits_{K^+} dh(n)\mu(n)\ .
\label{int}
\end{equation}
The measure has the form  \cite{s1}
\begin{equation}
\mu(n) = \prod\limits_{\alpha >0}(h(n),\alpha)^2\ ,\label{measure}
\end{equation}
where $\alpha$ ranges all positive roots of the Lie algebra $X$.
The Cartan subalgebra is isomorphic to an $r$-dimensional Euclidean
space. The invariant scalar product can be thought as an ordinary
vector scalar product in it. Relative orientations and norms of
the Lie algebra roots are determined by the Cartan matrix \cite{hel}.
The integration measure for the other fields remains unchanged.

For example, $G= SU(2)$, then $r=1$, $\mu =h^2(n)$ where $h(n)$ is
a real number because $H_{SU(2)}$ is isomorphic to a real axis.
The Weyl chamber is formed by positive $h(n)$. The $su(3)$ algebra
has two simple roots $\omega_{1,2}$ ($r=2$). Their relative orientation
is determined by the Cartan matrix, $(\omega_1,\omega_2) = -1/2,\
|\omega_{1,2}|= 1$. The Weyl chamber is a sector on a plane (being
isomorphic to $H_{SU(3)}$) with the angle $\pi/3$. The algebra has
three positive roots $\omega_{1,2}$ and $\omega_1 +\omega_2$. So,
the measure (\ref{measure}) is a polynom of the sixth order.
Its explicit form is given by (\ref{su3}).

The field $h(n)$ is invariant under Abelian gauge transformations
\begin{equation}
g_H(n)h(n)g^{-1}_H(n) = h(n),\ \ \ g_H(n)\in G_H\ . \label{h}
\end{equation}
Therefore, after integrating out the coset variables $\tilde{g}(n)$
in accordance with (\ref{int}), we represent the partition function
of Yang-Mills theory as a partition function of the effective
Abelian gauge theory
\begin{equation}
Z_{YM}(\beta) = (2\pi)^{-Lr}\int{\cal D}\tilde{U}_\mu e^{-\beta S_W}
F(\tilde{U})\ ,
\label{ymza}
\end{equation}
where
\begin{eqnarray}
F(\tilde{U}) &= & (\det\beta D_\mu^TD_\mu)^{1/2}\int\limits_{K^+}
\prod\limits_{n} \left(dh(n)\mu(n)\right) e^{-\beta S_H}\ , \label{f(u)}\\
S_H &= & 1/2\sum\limits_{n,\mu}
tr\ \left(h(n+\hat\mu) -\tilde{U}_\mu^{-1}(n)h(n)
\tilde{U}_\mu(n)\right)^2\ . \label{sh}
\end{eqnarray}
To obtain (\ref{ymza}), we have done the integral over both the
Grassmann variables and the boson ghost field $\tilde{\varphi}(n)$,
which yields $(\det \beta D_\mu^TD_\mu)^{1/2}$.

The function
$F(\tilde{U})$ is invariant only with respect to Abelian gauge
transformations, $\tilde{U}_\mu(n)\rightarrow
g_H(n)\tilde{U}_\mu(n)g^{-1}_H(n+\hat\mu)$. It provides a dynamical
reduction of the full gauge group to its maximal Abelian subgroup.
Since no explicit gauge condition is imposed on the link
variables $\tilde{U}_\mu(n)$, the theory do not have
usual gauge fixing deceases,
like the Gribov copies or horizon. We shall call the Abelian
projection thus constructed a {\em dynamical Abelian projection}.

{\bf 3}. Making a coset decomposition of the link variables \cite{kronfeld}
\begin{equation}
\tilde{U}_\mu(n) =U^H_\mu(n) U^{ch}_\mu(n)\ , \label{cd}
\end{equation}
where $U^H_\mu(n) =\exp u^H_\mu(n),\ u^H_\mu(n) \in H$ and
$U^{ch}_\mu(n) =\exp u^{ch}_\mu(n),\ u^{ch}_\mu(n) \in X\ominus H$,
we conclude that lattice Yang-Mills theory is equivalent to
an Abelian gauge theory with the action
\begin{equation}
S_A = S_W -\beta^{-1}\ln F\ . \label{aa}
\end{equation}
The link variables $U^{ch}_\mu(n)$ play the role of charged fields,
while $U^H_\mu(n)$ represents "electromagnetic" fields. In the naive
continuum limit, $U^H_\mu$ become Abelian potentials
\begin{equation}
U^H_\mu(n) \rightarrow \exp \int\limits_{n}^{n+\hat\mu}dx^\mu A_\mu^H
\ , \ \ \ \  A_\mu^H\in H \ .
\label{ap}
\end{equation}
Note that the field $h(n)$ carries no Abelian charge and does not
interact with $U^H_\mu$ as easily seen from (\ref{cd}) and (\ref{sh})
because $(U_\mu^H)^{-1}(n)h(n)U_\mu^H(n) =h(n)$.

Bearing in mind results on simulations of the Polyakov loop dynamics
on the lattice, one should expect that the Coulomb field of charges
in the effective Abelian theory is squeezed into  stable tubes connecting
opposite charges. A mechanism of the squeezing has to be found from
a study of dynamics generated by (\ref{aa}). First, one should verify
if the dual mechanism can occur in the effective Abelian
theory.

In our approach, configurations $U_\mu^H(n)$ containing
monopoles can exist. Kinematical arguments
for this conjecture are rather simple.
Let $G$ be SU(N). In a matrix representation,
the change of variables (\ref{cvphi}) becomes
singular at lattice sites where the field $\phi(n)$ has
two coinciding eigenvalues. This condition implies
three independent conditions on components of $\phi(n)$ which
can be thought as equations for the singular sites.
At each moment of lattice time, these three equations
determine a set of spatial
lattice vertices (locations of monopoles). Therefore
on a four-dimensional lattice, the singular sites form
world-lines which are identified with world-lines
of monopoles \cite{thooft}.
The new link variables
\begin{equation}
\tilde{U}_\mu(n) = \tilde{g}(n)U_\mu(n)\tilde{g}^{-1}(n+\hat\mu)
\label{mon}
\end{equation}
inquires monopole singularities via $\tilde{g}(n)$. Their density
can be determined along the lines given in \cite{kronfeld}.

So, monopole dynamics is the dynamics of
configurations $\phi(n)$ with two equal eigenvalues
in the full theory (\ref{ymz2}). If such configurations
are
dynamically preferable, then one can expect that
in the dynamical Abelian projection,
effective monopoles and antimonopoles form a condensate.

All monopole-creating configurations of the scalar field
$\phi(n)$ can easily be classified in a gauge invariant way.
First of all we observe that the change of
variables (\ref{cvphi}) is singular if its Jacobian vanishes
\begin{equation}
\prod\limits_n\mu(n) =0\ . \label{sin}
\end{equation}
We have to classify all configurations $\phi(n)$ which lead to
$\mu(n)=0$. The polynom (\ref{measure}) is invariant with
respect to the Weyl group. According to a theorem of Chevalley \cite{hel},
any polynom in $H$ invariant with respect to $W$ is a polynom
of basis (elementary) invariant polynoms $tr\ h^l(n)$ with $l =
l_1, l_2,..., l_r$ being the orders of independent Casimir operators
of $G$ \cite{hel}. Therefore,
\begin{eqnarray}
\mu(n) &= & P(tr\ h^{l_1}(n), tr\ h^{l_2}(n)
, ..., tr\ h^{l_r}(n)) = \nonumber\\
 &= & P(tr\ \phi^{l_1}(n), tr\ \phi^{l_2}(n), ..., tr\ \phi^{l_r}(n)) =0\ .
 \label{sin2}
\end{eqnarray}
Solutions of this algebraic equation determine all configurations $\phi(n)$
which will
create monopoles in the dynamical Abelian projection (\ref{ymza}).
For $G=SU(3)$, we have $r=2,\ l_1=2,\ l_2=3$ and \cite{s2}
\begin{equation}
\mu_{su(3)}(n) = \frac 12\left(tr\ \phi^2(n)\right)^3 -
3\left(tr\ \phi^3(n)\right)^2 =0\ .
\label{su3}
\end{equation}
Note also that $\mu_{su(3)}\sim (\phi_1-\phi_2)^2(\phi_2-\phi_3)^2
(\phi_3-\phi_1)^2$ where $\phi_{1,2,3}$ are eigenvalues of the
hermitian $3\times 3$ matrix $\phi\in su(3)$.

A dynamical question is: whether such
configurations are dynamically preferable in the full theory (\ref{ymz2}).
If they are not, the squeezing of the electrical field cannot
be explained by the dual mechanism because a creation of monopole-like
excitations would be dynamically unfavorable.  Studies
 of relative mean-values of gauge-invariant local
operators like $tr\ \phi^k(n)$
and of $\mu(n) =P$ in the full theory (\ref{ymz2}) could answer this
question. Since (\ref{sin2}) determines all configurations of $\phi(n)$
which could create topological monopole-like excitations
in the Abelian theory, the above investigation of dynamics would also show
if these effective excitations are indeed relevant to the squeezing the
Abelian electrical field and, hence, to the QCD confinement. Clearly,
the approach is gauge invariant.

{\bf 4}. Dynamics of monopoles is described by configurations
of an auxiliary field $\phi(n)$
satisfying the gauge invariant condition (\ref{sin2}). As
the field $\phi(n)$ is
coupled to
gauge fields in a standard (gauged) way,
it is natural to find configurations
of the link variables in the full theory (\ref{ymz2}) which
turns into monopole-carrying configurations of $U_\mu^H$ in
the dynamical Abelian projection. These configurations must
be relevant for the confinement, provided the dual mechanism
does occur in the dynamical Abelian projection.

As follows from (\ref{sh}) and (\ref{cd}), the Abelian
field $U_\mu^H(n)$ and the Cartan field $h(n)$ are decoupled
because $[U_\mu^H(n), h(n)] =0$. So, in the full theory,
we define Abelian link variables by the relation
\begin{equation}
[U_\mu^\phi(n), \phi(n)] =0\ .
\label{al}
\end{equation}
The coset decomposition assumes the form
\begin{equation}
U_\mu(n)=U_\mu^\phi(n)U_\mu^{ch}(n)\ .
\label{cd2}
\end{equation}
One can regard it as a definition of charged fields $U_\mu^{ch}(n)$
for given $U_\mu(n)$ and $\phi(n)$.

Consider a vector potential corresponding to $U_\mu^\phi(n)$ as
determined by (\ref{ap}). It has the form
\begin{equation}
A_\mu^\phi(n) = \sum\limits_{\alpha=1}^r B_\mu^\alpha(n) e_\alpha^\phi(n)\ ,
\label{ap2}
\end{equation}
where $B_\mu^\alpha(n)$ are real numbers, and Lie algebra elements
$e_\alpha^\phi(n)$ form a basis in the Cartan subalgebra constructed
in the following way
\begin{equation}
e_\alpha^\phi =\lambda_i tr\ \lambda_i\phi^{l_\alpha-1} \ .
\label{hb}
\end{equation}
It is not hard to be convinced that \cite{s2}
\begin{equation}
[e_\alpha^\phi,\ e_\beta^\phi] =0\ .
\end{equation}
Since for any group $G$ one of the numbers $l_\alpha$ is equal to 2,
one of the elements (\ref{hb}) coincides with $\phi$ itself.
The elements (\ref{hb}) are linearly independent in $X$ because
\begin{equation}
\det P_{\alpha\beta}\equiv \det tr\ e_\alpha^\phi e_\beta^\phi
=const\cdot P\ .
\label{p}
\end{equation}
So, a generic element $\phi$ of $X$ has a stationary group
$G_\phi\subset G$ with respect to the adjoint action of $G$
in $X$, $g_\phi\phi g_\phi^{-1} =\phi,\ g_\phi\in G_\phi$.
This stationary group is isomorphic to the Cartan subgroup
$G_H$. All linear combinations of the elements (\ref{hb})
form a Lie algebra of $G_\phi\sim G_H$.

In fact, the basis (\ref{hb}) can be constructed without
an explicit matrix representation of $\lambda_i$. We recall
that for each compact simple group $G$ and its Lie algebra $X$,
there exist $r={\rm rank}\ G=\dim H$ symmetrical irreducible tensors of
ranks $l_\alpha$, $d_{i_1,i_2,...,i_{l_\alpha}}$, invariant with
respect to the adjoint action of $G$ in $X$. Clearly,
$(e_\alpha^\phi)_i = d_{ij_1...j_{l_\alpha-1}} \phi_{j_1}\cdots
\phi_{j_{l_\alpha-1}}$.

Now it is easy to see that the Abelian potentials $B_\mu^\phi(n)$
are singular at lattice sites where $\phi(n)$ satisfies (\ref{sin2}).
Indeed, from (\ref{ap2}) we get
\begin{equation}
B^\alpha_\mu(n) = P^{\alpha\beta}(n)tr\ e_\beta^\phi(n) A^\phi_\mu(n)\ ,
\label{b}
\end{equation}
where $P^{\alpha\beta}P_{\beta\gamma}=\delta^\alpha_\gamma$.
The determinant of the matrix $P_{\alpha\beta}(n)$ vanishes
at the sites where $\mu(n)=P(n)=0$. At these sites, the
inverse matrix $P^{\alpha\beta}(n)$ does not exist, and the
fields $B_\mu^\alpha(n)$ are singular. For unitary groups
SU(N), $l_\alpha =2,3,...,N$, the singular sites form
lines in the four-dimensional lattice \cite{thooft},\cite{kronfeld}.
These lines are world-lines of monopoles.

{\bf 5}. The above procedure of avoiding explicit gauge fixing
can be implemented to remove the gauge arbitrariness completely
and, therefore to study gauge-variant correlators,
like the gluon propagator, or some other quantities requiring
gauge fixing on the lattice \cite{phde}. The advantage of dynamical gauge
fixing is that it is free of all usual gauge fixing dynamical
artifacts, Gribov's ambiguities and horizon \cite{scholtz}. It is
also Lorentz covariant.

Recent numerical studies of the gluon propagator in the
Coulomb gauge \cite{bernard}  show that it can be fit to a
continuum formula proposed by Gribov \cite{gribov}.
The same predictions were also obtained
in the study of Schwinger-Dyson equations where no effects of
the Gribov horizon have been accounted for \cite{stingl}.
The numerical result does not exclude also a simple massive
boson propagator for gluons \cite{bernard}. So, the problem
requires a further investigation.

Gauge fixing singularities (the Gribov horizon)
occur when one parametrizes the topologically nontrivial
gauge orbit space by Cartesian coordinates. So, these singularities
are pure kinematical and depend on the parametrization (or gauge) choice.
They may, however, have a dynamical
evidence in a gauge-fixed theory \cite{mitr}.
For example, a mass scale determining a nonperturbative pole
structure of the gluon propagator in the infrared region
(gluon confinement) arises from the Gribov horizon
\cite{gribov}, \cite{zwanziger} if the Lorentz (or Coulomb) gauge
is used. From the other hand, no physical quantity can depend on
a gauge chosen. There is no gauge-invariant
interpretation (or it has not been found yet) of the above mass scale.
That is what makes the gluon confinement model based on the
Gribov horizon looking unsatisfactory.

Here we suggest a complete dynamical reduction of the gauge symmetry
in lattice QCD, which involves no gauge condition imposed on
gauge fields and, hence, is free of
the corresponding kinematical artifacts.

For the sake of simplicity, we discuss first the gauge group SU(2).
Consider two auxiliary (ghost) complex fields $\psi$ and $\phi$,
Grassmann and boson ones, respectively. Let they realize the
fundamental representation of SU(2), i.e. they are isotopic spinors.
The identity (\ref{id}) assumes the form
\begin{equation}
Z_b(\beta)Z_f(\beta) = \int {\cal D}\phi^+{\cal D}\phi
{\cal D}\psi^+{\cal D}\psi e^{-\beta(S_b +S_f)} = 1\ ,
\label{id2}
\end{equation}
where $S_f = \sum_n (\nabla_\mu\psi)^+\nabla_\mu\psi$ and
$S_b = 1/2\sum_n (\nabla_\mu\phi)^+\nabla_\mu\phi$, and the
lattice covariant derivative in the fundamental
representation is defined by $\nabla_\mu\phi(n)
=\phi(n+\hat\mu) -U_\mu^{-1}(n)\phi(n)$. Inserting the identity
(\ref{id2}) into the integral representation of the Yang-Mills
partition function (\ref{ymz}), we obtain an effective gauge
invariant action. The ghost fields are transformed as
$\phi(n)\rightarrow g(n)\phi(n)$ and $\psi(n)\rightarrow g(n)\psi(n)$.

In the integral (\ref{ymz2}), we go over to new variables to
integrate out the gauge group volume
\begin{equation}
\int d\phi^+(n) d\phi(n) = v_{su(2)}\int\limits_0^\infty d\rho(n)
\rho^3(n)\ , \label{int2}
\end{equation}
where $\phi(n)= \tilde{g}(n)\chi\rho(n),\ \chi^+=(1\ 0)$, $\rho(n)$
is a real scalar field,  and $\tilde{g}(n)$ is a generic element
of SU(2). A new fermion ghost field and link variables $\tilde{U}_\mu$
are related to the old ones
via a gauge transformation with $g(n) =\tilde{g}^{-1}(n)$. Since
the effective action is gauge invariant, the integral over
$\tilde{g}(n)$ yields the gauge group volume $v_{su(2)}^L$.
We end up with the effective theory
\begin{eqnarray}
Z_{YM}(\beta) & =& \int{\cal D}\tilde{U}_\mu e^{-\beta S_W}F(\tilde{U})\ ,
\label{43}\\
F(\tilde{U}) &= &(\det \beta\nabla_\mu^+\nabla_\mu)^{1/2}
\int\limits_{0}^{\infty}\prod_n
\left(d\rho(n)\rho^3(n)\right) e^{-\beta S(\rho)}\ ,
\label{44}\\
S(\rho)&=&1/2\sum\limits_{n,\mu}\left(\rho(n+\hat\mu) -
\chi^+\tilde{U}_\mu^{-1}(n)\chi\rho(n)\right)^2\ .
\label{45}
\end{eqnarray}
The function (\ref{44}) is not gauge invariant and provides the dynamical
reduction of the SU(2) gauge symmetry. A formal continuum theory
corresponding to (\ref{43}) has been proposed and discussed in
\cite{scholtz}.

Expectation values of a gauge-variant quantity $G(U)$ are determined by
\begin{equation}
\langle G(U)\rangle \equiv \langle F(U)G(U)\rangle_W =
\int {\cal D}U_\mu e^{-\beta S_W}F(U)G(U)\ .
\label{46}
\end{equation}
For example, for the gluon two-point correlator one sets
$G(U) =A_\mu(n)A_{\mu'}(n')$ where the gluon vector potential
on the lattice reads
\begin{equation}
2ia A_\mu(n) = U_\mu(n) -U^+_\mu(n) -\frac 12 tr\ (U_\mu(n) - U^+_\mu(n))
\label{a}
\end{equation}
with $a$ being the lattice spacing.

For gauge groups of higher ranks, like SU(N), the dynamical
reduction can be done in a few steps as suggested in \cite{scholtz}. Note
that the procedure (\ref{id2})--(\ref{45}) being applied to
SU(N) would reduce the gauge symmetry to SU(N-1). To reduce
the gauge symmetry completely, one should repeat the procedure
$N-1$ times.

Another simpler way is to start with the dynamical Abelian
projection (\ref{ymza}). The Abelian $G_H$-symmetry can be
reduced in a way similar to (\ref{id2})--(\ref{45}), or
one can just impose the Lorentz gauge on the Abelian potentails
(link variables). The latter procedure is exempt of Gribov's copying.
Thus,
in this approach, a dynamical reduction of any gauge
symmetry group can be done in two steps.

Due to a complicated function (\ref{44}) involved in (\ref{46}),
numerical simulations are harder to carry out  than for
a usual gauge fixing procedure. However,
they could shed a light on the origin of the nonperturbative
pole structure of the gluon propagator. In this approach, the
gauge-dependent influence of the gauge fixing singularities,
like Gribov horizon, on the gluon propagator poles is excluded.

A relation of (\ref{ymza}) and (\ref{43}) to the corresponding
path integral representation of them with usual gauge fixing
on the lattice can be established by making a change of variables.
If one has to transform the integral (\ref{43}) to the integral
in a gauge $f(U)=0$, the change of variables should have the form
$\tilde{U}_\mu(n) = g(n)U^f_\mu(n) g^{-1}(n+\hat{\mu}),\
\phi(n) = g(n)\chi \rho(n)$ where $f(U^f)\equiv 0$.
The measure reads $\int{\cal D}\tilde{U}_\mu\int_0^\infty
{\cal D}\rho\prod_n \rho^3(n) = \int_{\Lambda_f}{\cal D}U_\mu^f
\Delta_{FP}(U^f)\int {\cal D}\phi$ where $\Delta_{FP}$ is the
Faddeev-Popov determinant in the gauge chosen and $\Lambda_f$ is
the fundamental modular domain for it. The integral over $\phi$
can be done and cancels the determinant in (\ref{44}). So, the
integral (\ref{43}) assumes a usual gauge-fixed form. This establishes
also a relation of the above approach with a standard gauge-fixed
perturbation theory.

{\bf 6}. An extension of our approach to the full lattice QCD does not
meet any difficulty because quarks would be decoupled with
the ghost fields.

Though the integration domain is restricted in the sliced path
integral (\ref{f(u)}), this restriction will disappear in the
continuum limit because of contributions of trajectories
reflected from the boundary $\partial K^+$ \cite{s1}, \cite{s2}.
It is rather typical for gauge theories that a scalar product
for physical states involves an integration over a domain
with boundaries which is embedded into an appropriate Euclidean
space. The domain can even be compact as, for example, in two-dimensional
QCD \cite{s3}. In the path integral formulation, this feature
of the operator formalism is accounted for by appropriate boundary
conditions for the transition amplitude (or the transfer matrix)
rather than by restricting the integration domain in the
corresponding path integral \cite{s3}, \cite{s4}. In turn,
the boundary conditions are to be found from the operator
formulation of quantum gauge theory \cite{s2}, \cite{s3}, \cite{s4}.
So, a study of the continuum limit requires an operator
formulation of the dynamical reduction of a gauge symmetry, which
has been done in \cite{scholtz2}.

The dynamical Abelian projection can be fulfilled in the continuum
operator formalism. The whole discussion of monopole-like singular
excitations given in sections 3 and 4 can be extended to the continuum
theory. So, it determines Lorentz covariant dynamics of monopoles
free of gauge fixing artifacts. To study monopole dynamics in the
continuum Abelian gauge theory, one has to introduce monopole-carrying
gauge fields \cite{kleinert}.

\begin{center}
 {\bf Acknowledgement}
\end{center}

I express my gratitude to F. Scholtz for valuable discussions
on dynamical gauge fixing, to A. Billoire, A. Morel
and V. K. Mitrjushkin
for providing useful insights about lattice simulations,
and D. Zwanziger and
M. Schaden for a fruitful discussion on the Gribov problem.
I would like to thank J. Zinn-Justin for useful comments on
a dynamical evidence of configuration space topology
in quantum field theory. I am very grateful to H. Kleinert for
a stimulating discussion on monopole dynamics.

%\newpage

\end{document}